\documentclass{article}
\usepackage{enumerate}
\usepackage{url}
\usepackage{mathtools}
\usepackage{amsthm}
\usepackage{amsfonts}
\usepackage{listings}

\makeatletter
\def\bibtex@style{amsrn-mod}
\makeatother

\RequirePackage{helvet,cclicenses}
\PassOptionsToPackage{shortlabels}{enumitem}
\PassOptionsToPackage{inline}{enumitem}
\RequirePackage{enumitem}
\setlist[enumerate]{font=\normalfont,labelindent=*,leftmargin=*,start=1}
\setlist[itemize]{labelindent=*,leftmargin=*}
\setlist[description]{labelindent=*,leftmargin=*,itemindent=-1 em}
\RequirePackage{microtype}
\RequirePackage{tikz}
\RequirePackage{color}
\RequirePackage[hidelinks]{hyperref}
\RequirePackage{etoolbox}

\newtheorem{thm}{Theorem}[section]

\newcommand*{\email}[1]{%
    \normalsize\href{mailto:#1}{#1}\par
    }

\newtheorem{definition}{Definition}

\newcommand*{\N}{\mathbb{N}}
\newcommand*{\R}{\mathbb{R}}
\newcommand*{\Z}{\mathbb{Z}}

\newcommand*{\J}{\mathbb{J}}

\providecommand{\keywords}[1]{\textbf{\textit{Keywords: }} #1}

\def\endfront@text{}

\usepackage{listings}
\usepackage{color}

\definecolor{dkgreen}{rgb}{0,0.6,0}
\definecolor{gray}{rgb}{0.5,0.5,0.5}
\definecolor{mauve}{rgb}{0.58,0,0.82}
\DeclarePairedDelimiter\abs{\lvert}{\rvert}%

\begin{document} 

\markboth{Authors' Names}
{Instructions for Typing Manuscripts $($Paper's Title\/$)$}
  
\title{Epi-constructivism: Decidable sets of computable numbers as foundational objects for mathematics}
\maketitle

\author{Zvi Schreiber}
\email{zvi@zvi.net}

\begin{abstract}
It is well known that the $\R$, the set of real numbers, is an abstract set, where \textit{almost all} its elements cannot be described in any finite language. 

We investigate possible approaches to what might be called an \textit{epi-constructionist} approach to mathematics. While most constructive mathematics is concerned with constructive proofs, the agenda here is that the objects that we study, specifically the class of numbers that we study, should be an enumerable set of finite symbol strings. These might also be called \textit{decidable constructive real numbers}, that is our class of numbers should be a computable sets of explicitly represented computable numbers.

There have been various investigations of the \textit{computable numbers} going back to Turing. Most are however not expressed constructively, rather \textit{computable} is a property assigned to some of the abstract real numbers. Other definitions define constructive real numbers without reference to the abstract $\R$, but the construction is undecidable, i.e., we cannot determine if a given construction represents a computable real number or not. For example, we may define a real as a computable convergent sequence of rationals, but cannot in general \textit{decide} if a given computable sequence is convergent.

This paper explores several specific classes of decidable constructive real numbers that could form foundational objects for what we might call an \textit{epi-constructionist} mathematics. These include \textit{Primitive Recursive Computed Numbers} namely a binary power series where the coefficients are provided by a primitive recursive function, the broader and most useful class of \textit{Signed Primitive Recursive Computed Numbers}, and the broader still \textit{Generalized Computed Numbers} which even include the limit of the Specker Sequence which is not a computable number in the traditional sense.
\end{abstract}


\section{Foreword}
As an informal introduction, let us consider the riddle that a friend told me over dinner which motivated this line of research. Versions of the riddle are found in \cite{hardin2008}.

An infinite number of people numbered $1, 2, 3,...$ are given time to confer, then they are arranged in a line so that person $n=k$ is facing all the people with $n>k$. Each has a hat with a random natural number placed on their head (the distribution is not important). Each must guess the number on their head, with all but a finite number guessing correctly.

This challenge appears impossible, given that nobody has any information at all about the hat on their head and so they are purely guessing. But the following solution is based on a suggestion attributed in \cite{hardin2008} to Yuval Gabay and Michael O'Connor, as graduate students at Cornell University in 2004. 

While confering, the people consider all sequences $(a_i)$ of natural numbers grouped into the equivalence class of sequences which defer at only finitely many positions. The axiom of choise ensures that there is a choice of sequences which includes just one representative of each equivalence class. Imagine the people could make such a choice.

Now once lined up with the hats, each person can see all but finitely many hats, and can therefore determine the equivalence class of the full sequence of hats. They assume that the hats are in fact arranged according to the representative sequence of that equivalence class, and guess their own hat accordingly. Since the real sequence of hats, and the class representative defer at finitely manly places, all but finitely many people guess their own hat correctly.

Of all the paradoxical corollaries of the axiom of choice, it was this one that I find hardest to accept. Clearly while the axoim of choice stiuplates that there exists a choice of a representative of each equivalence class of sequences, this choice could not in fact be constructed. So in what sense does it exist? This riddle motivated a study of whether we could have a foundation of mathematics which only reasons about objects which we can actually represent.

\section{Introduction}

Although we can discuss the axioms of the real numbers $\R$ as an abstract set, it is well known that there is limited opportunity to describe the specific elements of the set, as only countably many elements of $\R$ can be described by any finite expression in any natural or mathematical language; consequently, \textit{almost all} elements of $\R$ are inherently indescribable. Related to this, it is well known that there is not a unique model of the axioms of $\R$ (there is not even a unique model of the Peano axioms \cite{peano1889arithmetices} of $\N$), and there are some fundamental undecidable questions regarding the nature of $\R$; for example, the continuum hypothesis that $\R$ doesn't have a subset strictly larger than $\N$ is undecidable \cite{Cohen1966}.

There some well-known constructive approaches to mathematics. 
Most work in constructive mathematics focuses on the nature of mathematical proof, namely the banning of proving $\exists x:P(x)$ without finding $x$. With this limitation, the axiom of choice is discarded, or severely limited, and even the simplest choice, the law of excluded middle (LEM), is dropped from logic. Thus, we cannot prove $p \vee q$ without providing at least one of $p$ or $q$. This approach is associated mostly with Brouwer's intuitionism \cite{Brouwer1923}. 

Historically, it came as a surprise to many when Bishop showed that a large portion of mathematics survives this limitation \cite{bishop1967foundations}. Constructive expositions of real analysis have continued to the present  \cite{aberth2001computable,Bridger2006,BridgesVita2006,BridgesVita2019}.

Here we instead consider what we will call \textit{epi-constructivism}, so called to distinguish it from \textit{constructivism}. The principle is that the objects which we wish reason about--the numbers, and more generally the elements in our sets--are constructed explicitly, in the sense that we can write down every object using a string of symbols and that given a string of symbols we can determine in finite time if it is one of our objects or not. 

This requirement of constructive objects is orthogonal to the requirement of constructive proofs.
That is, we may choose to study only epi-constructive objects, and reject sets like $\R$ which have indescribable elements, but we do not necessarily reject LEM. Conversely, we may accept non-constructive proofs of $\exists x\in\R:P(x)$ without identifying which specific $x$ has the property, while however rejecting the abstract set $\R$ and focusing instead on classes of constructive objects, maintaining that even if one doesn't know which $x$ has the property, one knows that it is some describable $x$.

It was A. A. Markov who most clearly set an agenda that mathematics should study objects that may be explicitly constructed. Quoting from \cite{Markov1962,Markov1971translation}:
\begin{quote}
Constructive objects are certain figures that are put together in
particular ways from elementary figures, which are the elementary constructive
objects. An example would be the structures built up with the help of a
child's ``Erector'' set [like lego]\ldots In constructive mathematical
theories we limit ourselves to the consideration of constructive objects
of some standard type\ldots One of the simplest types of constructive objects are the words in a
particular fixed alphabet. A word in a given alphabet is a string of letters
of that alphabet\ldots
\end{quote}
Markov constructs the natural numbers, integers, rationals, and finally real numbers in such as way that they can be written as strings of symbols:
\begin{quote}
The natural numbers are introduced as words \texttt{0, 01, 011, \ldots} in the alphabet \texttt{\{0, 1\}}. The rational numbers are introduced as certain words in the alphabet of rational numbers \texttt{\{0, 1, -, /\}}\ldots\,
The normal algorithms in this alphabet that take each natural number into some rational number are called \textit{constructive sequences of rational numbers}\ldots\,
Let $\mathbb{U}$ be a constructive sequence of rational numbers. We will say that $\mathbb{U}$ \textit{converges regularly} if $\forall:i\leq j \implies i|\mathbb{U}(i)-\mathbb{U}(j)|<1)$\ldots

We will call \textit{proper real numbers} the words\ldots\, encoding
of a regularly convergent sequence. The \textit{constructive real numbers}\ldots\, will be the proper real numbers and the rational numbers.
\end{quote}

As paraphrased by \cite{Kushner2006}:
\begin{quote}
A pair of algorithms is called a \textit{constructive real number} if the first algorithm is a constructive sequence of rational numbers and the second effectively estimates the rate of convergence of this sequence. For such constructive real numbers one can define in some natural way the relations of order and equality as well as arithmetical operations. 
\end{quote}

In other words, the second algorithm gives a \textit{modulus of Cauchy convergence} to ensure constructively that the first algorithm yields a convergent sequence of rational numbers.

A similar definition may be found, among other equivalent definitions, in \cite{rice1954recursive} 
\begin{quote}
A recursively enumerable sequence of rational numbers $a_0, a_1, a_2,\ldots$ is recursively convergent (r.c.) when there exists a general recursive function $g(x)$ such that $|a_n - a_m| < \frac{1}{N}$ for $g(N) <n, m$. And a \textit{recursive real number} is the limit of an r.e., r.c. sequence of rational numbers.
\end{quote}

Unfortunately, all the above formulations of \textit{constructive real numbers} or \textit{recursive real numbers} fall short of the goal of having decidable elements. Given a pair of algorithms (representing a rational sequence and an explicit Cauchy modulus as in \cite{Kushner2006}) we cannot decide algorithmically if the second algorithm is indeed a valid modulus of Cauchy convergence for the first, as famously, we cannot even decide if it halts \cite{Turing1937}. All of the other definitions suffer from the same limitation. They allow a constructive description of real numbers, but the construction is not decidable; given a string of symbols we cannot generally determine if it represents a real number, particularly whether the algorithm describes a Cauchy sequence of rational numbers. Interestingly, Markov defined the integers and rationals in a decidably constructive manner, but then accepted an undecidably constructive definition of the reals.

This paper offers a preliminary investigation of epi-constructivism, by exploring sets of representations of real numbers (i.e., explicitly constructed convergent sequences of rational numbers) that are decidable in that the ``numbers'' are represented symbolically (as in Markov's formulation) but with the additional property that the class is a decidable (/enumerable/computable) set, meaning that given a string of symbols, a terminating algorithm can determine whether it is in the set or not. This also implies that our classes of constructive numbers will be enumerable, i.e., an algorithm can output an enumeration of all the valid constructive real numbers in the classes that we present.

The classes we will consider include \textit{Primitive Recursive Computed Numbers (PRCN)} which are expressed as a binary power series, where the coefficients are output by a primitive recursive function, and the important generalization of \textit{Signed Primitive Recursive Computed Numbers (SPRCN)} where the binary power series may have negative coefficients. Finally we consider Generalized Computed Numbers (GCN) which are algorithms that may not have an explicit power series, but where nevertheless we can confirm by inspection that the algorithm outputs a convergent sequence of rationals. This latter broadest group includes the limit of the Specker Sequence which is not normally considered computable.

In exploring decidable constructive real numbers, we may borrow not only from ideas of constructive mathematics, but also from the closely related ideas of computable analysis. Accordingly, we first present a very short survey of useful definitions of \textit{computable numbers}, as they are closely related to constructive real numbers.

\subsection{Computability as a property of abstract real numbers} 

Alan Turing opened his well-known paper \cite{TURING1936} with 
\begin{quote}
The ``computable" numbers may be described briefly as the real numbers whose expressions as a decimal are calculable by finite means.
\end{quote}

So a real number has the property of being computable if there exists a Turing machine, or computer program, which will output the decimal expansion one digit at a time. Note that Turing does not have any explicit constructive agenda. He accepts the existence of real numbers in the abstract sense, and explores which of them have the property of being computable.

This definition arguably places too much importance on the decimal expansion. In practice, it is also inconvenient to approximate a real number by sequentially outputing its decimal expansion due to the so-called table maker's dilemma \cite{Lefevre1998} \cite{Kahan2004}. For example, it is necessary to compute an accuracy of $10^{-10}$ before determining even the very first digit of the number $0.999999999$ as it is so close to $1.0$). Another disadvantage is that given a computer program that has output some decimal digits, we cannot generally determine whether it will output more digits or loop forever, as a result of the halting problem. 

Conversely, one advantage of Turing's definition is that a decimal expansion is inherently a convergent series and given a computer program that outputs only decimal digits, there is no need for a modulus of convergence to ensure that the output digits represent a convergent sequence. We will use this approach in some of our definitions.

More modern definitions do not rely on the decimal expansion. According to \cite{ZhengRettinger2004}, $x\in\R$ \textit{is computable}, is equivalent to the following:
\begin{quote}
There is a computable sequence $(x_s)$ of rational numbers which converges to $x$ effectively in the sense that
$(\forall s,t\in\N)(t\geq s \implies |x_s-x_t|\leq 2^{-s}$).
\end{quote}
This paper, together with \cite{RETTINGER2006818} and other papers by the same authors, investigate an interesting hierarchy of narrower and broader classes of computable numbers, in the abstract sense of real numbers that are computable. The various definitions of computable numbers are equivalent to Turing's original definition \cite{robinson1951,rice1955,myhill1953}. All the definitions refer to real numbers, i.e., they assume the existence of the abstract concept of real numbers, and explore which real numbers have the property of being computable numbers. 

\subsection{Computable numbers, without reference to real numbers}
A definition that does not reference real numbers may be found in Aberth's book \cite{aberth2001computable} (his final definition 2.1) which defines a real number in the context of computable calculus, thus:
\begin{quote}
A [computable] real number $a$ is defined by a program $a(K)$ that
for any value of the positive integer $K$ supplies a rational approximant $m_K$ and
a nonnegative rational error bound $e_K$ defining an interval $m_K \pm e_K$ of rational
numbers. For any choice of integers $K_1$ and $K_2$, the intervals $a(K_1)$ and $a(K_2)$
intersect. Given any positive rational error bound $E$, it is certain that an integer $K_0$ exists
such that $e_K < E$ for $K > K_0$.
\end{quote}

Hence Aberth defines a computable number as a program that outputs a sequence of rational intervals where all of the intervals overlap, and the size of the intervals converges to zero. Here there is a convergence of the study of computable real numbers and constructive real numbers in that this definition of computable real numbers is constructive, providing concrete representations, and nowhere assuming the existence of real numbers in the abstract sense.

However, again the set of computable real numbers satisfying this definition is not decidable. We cannot determine if a program $a(K)$ has the given properties, as we cannot even determine if $a(K)$ terminates and outputs any numbers $m_K$, $e_K$.

Aberth then defines that the programs $a(K)$ and $b(K)$ represent the ``same computable number'' if the intervals $a(K_1)$ and $b(K_2)$ overlap for all $K_1$ and $K_2$.  However, there is no deterministic way to determine if programs $a(K)$ and $b(K)$ have this property (again, we cannot even determine if either one halts for a given value). Hence, the equivalence relation and resulting equivalence classes are not computable or constructively valid. Therefore these computable real numbers, although not referencing the abstract reals, remain abstract in the sense that we cannot always determine if a given program represents a computable number at all, and we cannot necessarily determine if two programs are in the same equivalence class.

Another such approach is found in \cite{Weihrauch2000} defining a ``name'' (i.e. a representation) of a real number $x\in\R$ to be:
\begin{quote}
a sequence $(I_0,I_1,I_2,\ldots)$ of closed intervals $[a;b]$ with rational endpoints $a<b$ such that $I_{n+1}\subset I_n$ for all $n\in\N$ and $\{x\}=\bigcap_{n\in\N}I_n$. 
\end{quote}
He goes on to write:
\begin{quote}
 We assume tacitly that intervals are encoded appropriately, such that, strictly speaking, a name of a real number is an infinite sequence of symbols.
\end{quote}

Another recent related definition from \cite{BridgesVita2019} is:
\begin{quote}
 Given two families of intervals $F$ and $G$, we say $F$ is \textit{consistent} with $G$ if each interval from the family $F$ intersects each interval from the family $G$\ldots\, 
 A \textit{consistent family} of intervals is one that is consistent with itself\ldots\,
 The family $F$ is \textit{fine} when, for each positive rational $\epsilon$, there is an interval in $F$ that has length less than $\epsilon$\ldots\, 
 A [constructive] \textit{real number} is a fine and consistent family of rational intervals. 
\end{quote}
Two real numbers are defined to be equal if they are \textit{consistent} as families of intervals.

Hence, even the most modern approaches to constructive/computable real numbers do not provide decidably constructive real numbers.  Further, most authors attempted to define a non-determinable equality relation or equivalence class of the computable/constructive numbers, in an effort to recover the abstract concept of a real number, but at the expense of losing any kind of decidability of the equivalence classes.

\section{Seeking a computable set of computable numbers (recursive set of constructive real numbers)} 
A set is \textit{computable} (also known as recursive or decidable) if its characteristic function (which maps  elements of the set to $1$ and non-elements to $0$) is computable, that is, if a computer program can determine in a finite time whether a given element belongs to the set \cite{Davis1982}.

The term \textit{computable number} is variously use in the literature to refer to specific representations (i.e. the algorithm) or to the underlying abstract real number.
Henceforth we will use the term \textit{computed number} to emphasize that we are referring to the concrete representation (or ``name''), i.e., the algorithm or computer program that can compute a convergent sequence of rationals. The term computed number does refer to the abstract concept of a real number, which might be computable, nor to the concept from \cite{aberth2001computable} and others of an equivalence class of computed numbers, since these are non-determinable.

\begin{definition}
A  \textbf{Computed Number} is a syntactic representation of an algorithm (computer program) in some language that outputs a Cauchy sequence of rational numbers.
\end{definition}

This is not an entirely formal definition, as it leaves open the question of the language for algorithms, although per the Church-Turing thesis \cite{Church1936b}, this may be any Turing-complete language.

There are two desirable properties of a computed number. The first is that every sequence should be infinite, or should declare when it finishes, i.e., we don't prefer an algorithm which outputs a finite sequence of rational numbers and then fails to halt leaving us in suspense as to whether further rational numbers are forthcoming; rather the algorithm should preferably either output a finite sequence of rational numbers followed by a token representing ``finished'' (if the represented number is in fact rational) or should output a convergent infinite sequence. We call a computed number with this desirable property a \textit{Halting Computed Number}, in the sense that each time the program is called upon to produce the next number in the sequence it will halt (not in the sense that the overall sequence halts).

Another desirable property is that every computed number should have an explicit computable modulus of convergence, to enable us to know how long to run the algorithm to reach a given accuracy. We can call this a \textit{Modulated Computed Number}.
Without loss of generality, if we have a constructive method to output a convergent sequence of rationals $(q_i)$, we call it Modulated if it has a modulus of convergence $|q_i-q_j|<2^{-i} \, (j>i)$. In case there is a different modulus of convergence, we can simply adapt the algorithm to skip elements and yield, as the $i$th element, the first element for which the modulus of convergence gives $2^{-i}$.

Indeed for many constructive mathematicians, a sequence with an explicit modulus of convergence is the only acceptable Cauchy sequence (this is at least implied by the definition of a Cauchy sequence in \cite{bishop1967foundations}).

There are many (Modulated) (Halting) Computed Numbers in the literature quoted above. But our goal in this paper is to explore not individual numbers, but rather classes of computed numbers which are decidably so.

\begin{definition}
A  \textbf{Decidable class of (Modulated) (Halting) Computed Numbers} is a computable set of syntactic representations of algorithms (computer programs) in some language, (that is, it is decidable if a given algorithm representation is in the set), and where each algorithm in the set is a (Modulated) (Halting) Computed Number.
\end{definition}

Generally, we will present two useful approaches to defining a class of computed number algorithms for which we can decide deterministically if an algorithm belongs to the class. The first approach suitable for Modulated Halting Computed Numbers is insisting that the sequence has the form of an exponential power sequence $\sum_i P(i)2^{-i}$ where the $P(i)$ are in turn given by a well-defined algorithm, guaranteed to halt and output an integer within some bound. This type of Computed Number may be called a \textit{Power Series Computed Number}.

The second approach, which may produce computed numbers that are not halting or not modulated, is to take any computer program $P$ that outputs a sequence of rationals, and add a constraint to ensure that it outputs a convergent subsequence. We call this approach a \textit{Constrained Computed Number} and will return to it later.

\section{Primitive Recursive Computed Numbers (PRCNs)}
The most obvious approach to the goal of constructing a decidable class of Modulated Halting Computed Numbers is to represent numbers as a power series with bounded coefficients given by primitive recursive functions. Thus, the Primitive Recursive Computed Numbers are specified using an integer and then a fraction, specified with a power series with the coefficients given by a primitive recursive function. This ensures that the computation to any accuracy will terminate. A naive definition, inspired by Turing's original definition of a computable number, would be that the primitive recursive function takes an argument $n$ and returns $0$ or $1$ for the $n$th digit of the binary expansion. 
\begin{definition}
The class of \textbf{Primitive Recursive Computed Numbers (PRCN)} are the pairs $(I, P(n))$ where $I$ is an integer and $P:Z^+\to\{0,1\}$ is a primitive recursive function representing the rational series $I + \sum_{i=1}P(i)2^{-i}$ or equivalently the rational sequence $(I + \sum_{i=1}^n P(i)2^{-i})_n$.
\end{definition}

Primitive recursive functions are typically defined to be $P':\N\to\N$. But we can simply ignore $P'(0)$ and interpret any value $P'(i) \geq 1$ as a $1$m thereby giving us a function $P:\Z^+\to\{0,1\}$.

By limiting ourselves to primitive recursive functions (which have a decidable syntax and are always total), we can easily validate by means of a simple terminating algorithm whether any given $(I, P(n))$ is a PRCN. Hence, this definition realizes the goal of this paper of mathematical objects that are decidably constructive.

It should be clear that the PRCNs are a decidable class of Computed Numbers, indeed of Modulated Halting Computed Numbers. If we wish to be very strict regarding representing the elements as words in a language, à la Markov, we may write each PRCN as:
\begin{quote}
 $I$ \texttt{+} $\Sigma$ \texttt{|sgn(} $P(i)$ \texttt{)|*2}$\uparrow$\texttt{(-i)}
\end{quote}
where for $I$ we substitute any integer \texttt{O11}\ldots or  \texttt{-O11}\ldots and for $P(i)$ we substitute the syntactic representation of any primitive recursive function. Note the use of \texttt{|sgn()|} wherein \texttt{sgn} is the sign function and $||$ the absolute value, to constrain the value to $0$ or $1$. The summation implicitly ranges over $i=1,2,3\ldots$. A computer program could determine in finite time if a given string of symbols is in this class. 

Indeed, using what is now known as a \textit{Markov Algorithm} (which Markov called a \textit{Normal Algorithm}), we can build all the PRCNs with string rewriting rules, beginning with  \texttt{0+} $\Sigma$ \texttt{|sgn(i)|*2}$\uparrow$\texttt{(-i)} and applying the rules \texttt{0}$\rightarrow$\texttt{10},  \texttt{0}$\rightarrow$\texttt{-0} and rules to repeatedly substitute for \texttt{(i)} syntactic representations of the various primitive recursive functions: constant, successor, projection, composition, and primitive recursion.

Unfortunately, this definition leaves no flexibility. Every real number has a unique binary expansion (except for numbers of the form $I/2^n$ which have a finite expansion and classically also have a second representation with infinite trailing $1$'s).

Using this specific definition, it is therefore difficult to provide a Primitive Recursive Computed Number representing a real number as basic as say $\pi$. The reason again is the Table Maker's Dilemma, namely in order to calculate the $n$th digit in the binary expansion of $\pi$, it is necessary to calculate not an accuracy of $2^{-n}$ but rather an accuracy of $2^{-(n+k)}$ where $k$ is the number of consecutive $0$s or $1$s following the $n$th digit (or $9$'s in decimal). Given that $k$ is not known in advance, there is no obvious primitive recursive function to calculate the $n$th digit, as primitive recursive functions allow only loops with a pre-determined bounded number of iterations. 

(For $\pi$ specifically we can leverage the property $\abs{\pi-\frac{p}{q}}\geq q^{-42}$ \cite{KMahler1954} or better $\abs{\frac{\pi}{\sqrt{3}}-\frac{p}{q}}\geq q^{-4.6016}$ \cite{hata1992} to provide an upper bound on the number of consecutive $1$s (or $9$s) following the $n$th digit and use this bound to create a primitive recursive function which calculates the $n$th digit. But this would not be obvious a priori, and for other transcendental numbers there may not be such a bound.)

Hence the motivation to generalize the definition in the following section.
\section{Signed Primitive Recursive Computed Numbers (SPRCNs)}

\begin{definition}
The \textbf{Signed Primitive Recursive Computed Numbers (SPRCN)} are the pairs $(I, P(n))$ where $I$ is an integer and ${P}:\Z^+\to\{-1,0,1\}$ is a primitive recursive function representing the coefficients in the rational series
$I + \sum_{i=1}P(i)2^{-i}$, or equivalently the rational sequence $(I + \sum_{i=1}^n P(i)2^{-i})_n$. 
\end{definition}

Again if $P':\N\to\N$ is a primitive recursive function with the traditional domain and co-domain, we interpret the numbers $\{P(i)\}$ ($i>1$) as representing the coefficients $-1,0,1$ respectively depending if $P(i)>1$, $P(i)=0$, $P(i)=1$ respectively.
%

This generalized definition also gives us computed numbers as recognizable constructive objects in line with the goals of the paper. Syntactically we can write each SPRCN as:
\begin{quote}
 $I$ \texttt{+} $\Sigma^+$ \texttt{sgn(} $P(x)$ \texttt{)*2}$\uparrow$\texttt{(-i)}
\end{quote} 
where again $I$ is the syntatic representation of an integer and $P(x)$ a syntactic representation of a primitive recursive function, and we use the special summation $\Sigma^+$ which always takes the absolute value of the first non-zero summand if any.

Although in the abstract sense you could express any number using only the coefficients $0$ and $1$ as a PRCN, by allowing the $P(i)$ to have the value $-1$ we solve the Table Maker's Dilemma, as we can make corrections. For example if at first we have an estimate $2^{-1}$ or $0.1$ we can later correct it to $0.0111$ by subtracting $2^{-4}$. We will see in a theorem below that this enables us to express most known power series as SPRCNs.

We find SPRCNs to be the most natural and useful decidable class of Computed Numbers, indeed SPRCN is a Decidable class of Modulated Halting Computed Numbers.

It is a apparent that every Signed Primitive Computed Number converges to an abstract computable number in the classical sense, although we will see that the converse is not true. Of course, primitive recursive functions could be substituted with another class of total functions, although primitive recursive functions are widely considered to be the most natural class of functions that are obviously total.

Now we must prove that we can find Signed Primitive Computed Numbers to represent useful irrationals such as $\sqrt{2}$, $\pi$. 

For $\sqrt{2}$ we can easily formulate a signed primitive recursive function that initially guesses $2+\frac{1}{2}$ and then seeing that $(2+\frac{1}{2})^2>2$ guesses $2+\frac{1}{2}-\frac{1}{4}$, etc., fluctuating above and below $\sqrt{2}$ converging exponentially.

For $\pi$ we can use the Madhava–Leibniz sequence $\pi_n = 4\,-\,{\frac {4}{3}}\,+\,{\frac {4}{5}}\,-\,{\frac {4}{7}}\,+\,{\frac {4}{9}}\,-\,\cdots\pm\,{\frac {4}{2n-1}}$ which we can rewrite as $3+\frac{7}{15}\,\,\,-\,{\frac {4}{7}}+\,\frac {4}{9}\cdots$. 
Now we will show that in order to recursively choose $P(i)$ we need to find an $n$ such that 
$\forall m>n: |\pi_m-\pi_n| \leq 2^{-i-1}$. In this case $n=2^i$ would work. 
We then choose $P(i)$ so that $|(I  + \sum_{i=1} P(i)2^{-i}) - \pi_n|<2^{-i-1}$. 

(Of course in practice we may choose a faster converging series such as the Nilakantha series 
$\pi =3+{\frac {4}{2\times 3\times 4}}-{\frac {4}{4\times 5\times 6}}+{\frac {4}{6\times 7\times 8}}-{\frac {4}{8\times 9\times 10}}+\cdots $)

We will now formalize this and prove more generally that we have a Signed Primitive Recursive Computed Number (SPRCN) representation for any series of rationals with summands given by primitive recursive functions and with an explicit modulus of Cauchy convergence.

\begin{thm}\label{recursiveComputableSeriesTheorem}
 Let $(q_n)$ by any primitive recursive rational sequence in the sense that the $n$th element is given by $A(n)/B(n)$ where $A$ and $B$ are primitive recursive functions with $A$ any integer and $B$ positive. Let the sequence have a primitive recursive Cauchy modulus of convergence $C$ so that for every natural number $e$ representing the order of magnitude of a desired error, we have $\forall i,j \geq C(e): |q_i-q_j|<2^{-e}$). Then there is a Signed Primitive Recursive Computed Number with the same limit as $(q_n)$.
\end{thm}

\begin{proof}
Let $a_n = \sum_{i=0}^{\max(C(1)\ldots C(n+1))} A(n)/B(n)$ which is clearly a primitive recursive rational function of $n$ and defines a subsequence of $(q_n)$ which converges exponentially.
We show by induction that we can choose $I$ and $P(i) \in \{ -1, 0, 1\} \, (i=1\ldots n)$ to make $p_0 = I$ and $p_n = I + \sum_{i=1}^{n} P(i)\, 2^{-i} $ so that $|p_n - a_{n+1}| \leq 2^{-(n+1)}$. 

The base case of the induction is trivial by choosing the integer $I$ with $|I - a_1| \leq \frac{1}{2}$ which can be done by a primitive recursive function.

Now assuming $|p_{n-1} - a_n| \leq 2^{-n}$ and  we know by definition of the modulus of convergence $C$ that $|a_{n+1} - a_n| \leq 2^{-(n+1)}$ then by the triangle inequality $|p_{n-1} - a_{n+1}| \leq 2^{-(n+1)} + 2^{-n} = 3*2^{-(n+1)}$. Now a primitive recursive function can choose $P(n)= -1,0,1$ according to whether $a_{n+1} - p_{n-1}$ is in the interval $\left[-\frac{3}{2^{n+1}}, - \frac{1}{2^{n+1}}\right)$ or $\left[- \frac{1}{2^{n+1}},  \frac{1}{2^{n+1}}\right)$ or $\left[\frac{1}{2^{n+1}}, \frac{3}{2^{n+1}}\right) $ respectively, so that $a_{n+1} - p_{n} = a_{n+1} - p_{n-1} - P(n)/2^{n}$ will always be in the middle interval and will therefore satisfy $|p_{n} - a_{n+1}| \leq 2^{-(n+1)}$, completing the recursion.
\end{proof}

The SPRCN are our most natural decidable class of Computed Numbers. They are well behaved in that they always produce an infinite exponentially convergent sequence of rationals. Moreover as we have seen, the class includes representations of most popular irrationals like the algebraic numbers, $\pi$, $e$ etc. However, they cannot necessarily represent all abstract computable numbers, as our next result shows.

\subsection{PRCN and SPRCN are proper subsets of the computable numbers}
Let CN be the classical abstract set of computable real numbers as defined by Turing. Let $\overline{\text{SPRCN}}$ represent the abstract real numbers which may be represented by SPRCN computed numbers.

\begin{thm}
 $\overline{\text{SPRCN}} \subset \text{CN}$
\end{thm}

\begin{proof}
 $\overline{\text{SPRCN}} \subseteq \text{CN}$ is clear since every PRCN number can be computed by a computer program which sums the sequence.
 
 To find a CN number $N$ that is not represented by any SPRCN computed number we use the classical diagonal argument. Consider for simplicity only numbers between $0$ and $1$. Let $f_i$ be the $i$th primitive recursive function in a lexicographical enumeration of primitive recursive function, giving the power expansion of the SPRCN number $s_i$.
 
The approximation $\sum_{j=1}^{2}\frac{f_1(j)}{2^{-j}}$ gives $s_1$ to within $\pm\frac{1}{4}$. Hence, we can pick the first two digits of the binary expansion of $N$ (which fixed $N$ to an interval of width $\frac{1}{4}$) to ensure that it does not overlap with this interval. We can then calculate $s_2\approx \sum_{j=1}^{4}\frac{f_2(j)}{2^{-j}}$ to within $\frac{1}{16}$ and pick the next two binary digits of $N$ to ensure that $N$ will be in an interval of width $\pm\frac{1}{16}$ (within the interval of width $\frac{1}{4}$ which we have already fixed by the first two digits) not overlapping with the interval that contains $s_2$.  We continue in that manner to construct a computable number $N$ that cannot be equal to any SPRCN number $s_i$.
\end{proof}

\subsection{$\overline{\text{PRCN}}\subset\overline{\text{SPRCN}}$}
There is no direct means of converting an SPRCN computed number into a PRCN computed number.
An interesting example is the SPRCN number defined by $P(1)=1$, with $P(i)$ ($i>1)$ equal to $0$ if $2i$ can be expressed as the sum of two primes, otherwise $P(i)$ is $-1$ if $i$ is divisible by $13$ or $1$ otherwise.
Clearly this $P$ is a primitive recursive function. But we do not know even the first digit of the decimal expansion so cannot express it as a PRCN. Specifically, if the Goldbach Conjecture\cite{Goldbach1742} is true then the number is simply $0.1$ which can be expressed easily enough as a PRCN with either binary expansion $0.10000\dots$ or $0.01111\dots$. But in case the Goldbach Conjecture is false we cannot determine even the first digit of the binary expansion without knowing whether the smallest even number which cannot be expressed as the sum of two primes is divisible by $13$.

Accordingly, there is definitely no trivial transformation of SPRCNs to PRCN as solving the Goldbach Conjecture is not trivial, but that leaves open the question whether all abstract computable numbers that have a SPRCN representation also have a PRCN representation, or whether $\overline{\text{PRCN}}\subset\overline{\text{SPRCN}}$. Based on an idea from an anonymous reviewer whom I thank, we can construct an SPRCN number which has no possible PRCN representation.

\begin{thm}
$\overline{\text{PRCN}}\subset\overline{\text{SPRCN}}$
\end{thm}
\begin{proof}
It's known that the primitive recursive functions are recursively enumerable, and we start with enumeration $G_1, G_2, \dots$ of the primitive recursive functions $\N \rightarrow \{-1,1\}$ and define the nullary primitive recursive functions $g_i=G_i[i]$. We construct an SPRCN number $S$ which outputs a digit $1$, then ``executes" $g_1$ and outputs a binary digit of $0$ for every computation step of $g_1$ until it terminates, then outputs the final output of the program $|g_1|$. It then outputs another digit $1$, does the same for $g_2$. So we get an SPRCN number whose signed binary expansion is:
\begin{equation}
   S = 0\ .\ 1\ 0^{T[g_1]}\ |g_1|\ 1\ 0^{T[g_2]}\ |g_2| \ \dots
\end{equation}
where $T[g_i]$ is the number of computation steps till $g_i$ terminates and $0^{T[g_1]}$ means that many digits $0$ and $|g_i|$ is the output of program $g_i$.

To see that this is primitive recursive, i.e. computable using only bounded loops we calculate the $i$th digit of the signed binary expansion as follows:

\begin{lstlisting}
LOOP THROUGH THE PROGRAMS g_1... g_i FOR EACH:
   COMPUTE UP TO A MAX OF i STEPS OR UNTIL TERMINATED
   WHEN TOTAL NUMBER OF STEPS COMPUTED ACROSS THE PROGRAMS CONSIDERED SO FAR IS i:
      IF A PROGRAM IS IN MIDDLE OF EXECUTION HALT WITH OUTPUT 0
      ELSE HALT WITH OUTPUT OF THE PROGRAM WHICH JUST TERMINATED
\end{lstlisting}
The fact that a primitive recursive function is able to compute a bounded number of computation steps of the $n$th primitive recursive function is established in Kleene's recursion theorem.

However in the equivalent unsigned binary expansion, each block of digits is replaced e.g. $100001 \rightarrow 100001$ and $10000(-1) \rightarrow 011111$ so
\begin{equation}
S = 0\ .\ (\tfrac{1+|g_1|}{2})\ (\tfrac{1-|g_1|}{2})^{T[g_1]}\ 
1\ (\tfrac{1+|g_2|}{2})\ (\tfrac{1-|g_2|}{2})^{T[g_2]}\ 1\ \dots
\end{equation}

To see that this sequence is not primitive recursive, note that a function which outputs the $n$th digit must be able to determine the final output $|g_i|$. For example the first digit is $(\tfrac{1+|g_1|}{2})$. There is a well known argument by contradiction that a primitive recursive function cannot determine the final output of the $n$th primitive recursive function. Suppose $Q:\N \rightarrow \{-1,1\}$ is a primtiive recursive function such that $Q[i]=g_i=G_i[i]$. We can also construct $Q'[i]=-Q[i]$. Now $Q'$ must be in the enumeration so for some $j$ we have $Q' = G_j$. But then $Q'[j] = -G_j[j] = -Q'[j]$ providing the contradiction.
\end{proof}

\subsection{Equality of SPRCN numbers}
As is well known regarding computable numbers (e.g. \cite{aberth2001computable}), there is no general way to determine if two numbers are equal in the sense that the difference between their sequences converges to zero.

Signed Primitive Recursive Computed Numbers $(I_1,P_1(n))$ and $(I_2,P_2(n))$ could converge to the same abstract computable real number if $I_1=I_2$ and $\forall i: P_1(i)=P_2(i)$. They can also converge to the same abstract number without these conditions since we allow negative coefficients, the expansion is not unique. Even in the first case there is no general way to compute in finite time whether $\forall i: P_1(i)=P_2(i)$. So in general we cannot determine equality, although in some specific cases we would be able to.

This lack of equality is frustrating but reflects the reality that when we describe two real numbers we cannot always determine if we are describing the same number.

\subsection{Addition and subtraction of Signed Primitive Recursive Computed Numbers}
A corollary of Theorem \ref{recursiveComputableSeriesTheorem} is that we can follow the procedure in the proof to define addition of Signed Primitive Recursive Computed Numbers. For two Signed Primitive Recursive Computed Numbers $(I_1,P_1(n))$, $(I_2,P_2(n))$ we define the sum to be $(I_1 + I_2, (P_1+P_2)(n))$. Unfortunately we cannot add $(P_1+P_2)$ in the obvious pointwise manner as we want the parameters $p_i$ to always be $-1,0,1$.

So to compute \textbf{addition} $(P_1+P_2)$ we take the primitive recursive rational sequence $A(n)/B(n)$ created by adding $A(n) = P_1(n) + P_2(n)$ and $B(n) = 2^{n}$.  There is an obvious primitive recursive Cauchy modulus of convergence $C(e) = e+1$. Then we apply Theorem \ref{recursiveComputableSeriesTheorem}. 


Subtraction is achieved by adding the negation $(-I,-P)$.

\subsection{Multiplication of Signed Primitive Recursive Computed Numbers}
For \textbf{multiplication} of  two Signed Primitive Recursive Computed Numbers $(I_1,P_1(n))$, $(I_2,P_2(n))$ we can again define a primitive recursive sequence of rationals $A(n)/B(n) = (I_1 + \sum_{i=1}^n P_1(i) 2^{-i}) (I_2 + \sum_{i=1}^n P_2(i) 2^{-i})$ and define the primitive recursive Cauchy modulus of convergence $E(n) = n + \overline{log}(I_1+1) + \overline{log}(I_2+1)$ where $\overline{log}(x)$ gives $log_2(x)$ rounded up to an integer, and we again use Theorem \ref{recursiveComputableSeriesTheorem}.

These arithmetic definitions have the desired property that the traditional abstract limits will be added/multiplied in the classical sense. 

\subsection{Division of Signed Primitive Recursive Computed Numbers}
As is known in the field of computable analysis, it is not possible in general to calculate a multiplicative inverse, and we therefore cannot perform division. This is because given a computed sequence, we cannot determine in general if it converges to zero.

\section{Recursive Computed Numbers}
A natural extension (but one that turns out to be of limited use) is to drop the requirement for the function $P$ to be primitive recursive. Thus a \textbf{Recursive Computed Number} is a pair $(I, P(n))$ where $I$ is an integer and $P(n)$ is a recursive function which takes a positive integer and can only return the values $0, 1$ where $(I, P(n))$ represents the rational series $I + \sum_{i=1}P(i)2^{-i}$ or equivalently the rational sequence $(I + \sum_{i=1}^n P(i)2^{-i})_n$ which may terminate as a finite sequence in case some $P(m)$ doesn't terminate.

\begin{definition}
The \textbf{Recursive Computed Numbers} are the pairs $(I, P(n))$ where $I$ is an integer and $P(n)$ is a recursive function which takes a positive integer and which is syntactically structured to only ever return one of the values $0, 1$ where $(I, P(n))$ represents the rational sequence $(I + \sum_{i=1}^n P(i)2^{-i})_n$.
\end{definition}

In other words, in computing terms, we have an arbitrary computer program constructed to output a sequences of $0, 1$ providing the coefficients of a power series.
This definition is equivalent to that of the classical computable numbers, restated to be a Modulated Decidable class of Computed Numbers by focusing on the algorithm and removing any reference to the abstract real numbers. That is, we can determine by syntactic inspection if we have a recursive computed number.

However, clearly these recursive computed numbers are not in general Halting and therefore do not provide a satisfying foundation of constructive real numbers, as we may try to output successive rational approximations and be left hanging forever wondering if another approximation is forthcoming. 

We could again allow coefficients of $-1$, although a recursive function can calculated ahead an arbitrary number of places so is able to calculate power series without this relaxation.

%
%

\subsection{Arithmetic with Recursive Computed Numbers}
The limited usefulness of Recursive Computed Numbers becomes apparent when we try to do something as simple as adding two of them together, the challenge being that we never know if any one of the programs will ever output another number in the sequence.

Given two programs $P$, $Q$ it would be natural to alternately perform operations of programs $P$ and $Q$, and whenever there is a new number available from either output, the sum of the last two known numbers is output, so e.g. if the last outputs are $p_i$ and $q_j$ and then $q_{j+1}$ becomes available, we can output $p_i$ + $q_{j+1}$.  However, this sum is not a Recursive Computed Number, as we may for example reach $p_1 + q_{100}$ after which $p_2$ suddenly becomes available but we cannot output $p_2+q_{100}$ as this would involve a big jump out of turn.  We would obtain a better behavior if we output $p_2+q_2$ and then wait for $p_3+q_3$. This would be exponentially convergent but we can't do this either since we don't know and may never know whether $p_2$ actually exists, so we may get stuck with an empty output representing zero, which does not accurately represent the sum since $Q$ is non-zero.

Interestingly, the abstract computable numbers are a field, but there is no good way to perform arithmetic operations on the concrete representations.  That is if $p$ and $q$ are computable real numbers then we know there is a computer program which outputs a sequence converging to $p+q$ but there is no direct way to add programs $P$ and $Q$ to $P+Q$.

We will therefore not dwell on the Recursive Computed Numbers but rather move on to a more general class of Computed Numbers which has preferable properties.

\section{Generalized Computed Numbers}
\subsection{Naive GCNs}
It is well known that some Cauchy Sequences (indeed monotone bounded sequences) of rational numbers have limits that are not computable. We will investigate whether there are decidable classes of computed numbers which admit numbers that are not classically computable.

For convenience we  define a version of the \textbf{Specker sequence}  \cite{Specker49} $(S_i)$ as follows. Let $P_1, P_2,...$ be an enumeration of computer programs in some programming language,  lexicographically ordered.

Then define rational number $S_1$ as binary $0.1$ (i.e. $\frac{1}{2}$) if program $P_1$ halts after 1 step, $0.0$ otherwise.  
More generally $S_i$ is $0.b_{i1}b_{i2}...b_{ii}$ where $b_{ij}$ is a binary digit which is $1$ if $P_i$ halts in $j$ execution steps. Thus each $S_i$ adds one digit to the end of the binary expansion of $S_{i-1}$, but may also have some earlier digit flip from $0$ to $1$, that is the $j$th digit will flip from $0$ to $1$ if program $P_j$ $(j<i)$ terminates after exactly $i$ steps.

For example we might have
\begin{verbatim}
0.0
0.00
0.010
0.0101
0.01110
\end{verbatim}
where the $i$th digit flips from $0$ to $1$ in the $j$th number if program $P_i$ halts after $j$ steps.

Clearly this sequence is Cauchy in the abstract sense, as every digit in the binary expansion will either stay $0$ forever or flip to $1$ and then remain the same, so the sequence is monotone increasing with an upper bound of $0.1111\ldots=1$. But we cannot estimate it to a given error range, as we do not know in general if and when $P_1$ will terminate and the first digit might flip to $1$ creating a jump or $\frac{1}{2}$.

So the sequence is not Cauchy in the sense often used in constructive mathematics, since it does not have a computable modulus of Cauchy convergence.

However there is another means to constructively force a sequence to be Cauchy. If we can be sure that the sequence only has a specific finite number of jumps of each order of magnitude, then we can be certain it is Cauchy, even if we do not know exactly when it will reach a certain range of convergence.

This idea is presented in Rettinger and Zheng's \cite{RETTINGER2006818} definition of \textbf{dbc} computable numbers:
\begin{quote}
	We call an index pair $(i, j)$ a $2^{-n}$-jump if $|x_i - x_j| \geq 2^{-n}$. A sequence $(x_s)$ converges $f$-bounded effectively if it has at most $f(n)$ \textbf{non-overlapping} $2^{-n}$-jumps for all $n \in \N$.
	
	A real number $x$ is $f$-bounded computable (f-bc, for short) if there is a computable sequence $(x_S)$ of rational numbers which converges to $x$ f-bounded effectively. If a real number is $f$-bc for a computable function $f$, then it is also called \textit{divergence} bounded computable (\textbf{dbc}, for short)
\end{quote}

We might call the $f$ function a \textit{modulus of Cauchy divergence}.  Their idea is that even if we do not have a \textit{modulus of Cauchy convergence} i.e. we cannot predict when the sequence will settle down to a given error range, we can still ensure it's Cauchy if we instead limit the number of divergent jumps of every order of magnitude.

Clearly a sequence with such a \textit{modulus of Cauchy divergence} is Cauchy since for every $n$ there is a finite number of points $x_i$ such that there exists a point $x_j \, (j>i)$ with $|x_j-x_i|>2^{-n}$. Therefore, the tail of the sequence after the last such point satisfies the Cauchy condition.

Now \cite{RETTINGER2006818}'s definition is not a constructive definition as it depends on the abstract concept of a real number. Moreover, there is no deterministic way to decide if a given sequence represents a \textit{dbc} number. But this approach to defining a Cauchy sequence without a modulus of convergence but with an explicit bound on the number of jumps of each order of magnitude is useful in our constructive approach.

We will use the term \textbf{Generalized Computed Numbers (GCN)} for Unmodulated Decidable Classes of Computed Numbers which use a modulus of divergence instead of a modulus of convergence, so that they are known to converge conceptually but we do not know their rate of convergence.

\begin{definition}
	A naïve definition of \textbf{Generalized Computed Numbers (NGCN)} are the triples $(I, P(n,m), J(N))$ where $I$ is an integer and ${P}:\Z\times\Z\to\{0,1\}$ is a primitive recursive function with $P(n,m)$ giving the $m$'th version of the $n$th coefficient of the binary power series, while $J(n)$ is a primitive recursive function limiting how many times the $n$th coefficient $P(n,i)$ may change as $i$ increases. 
	
	The triple $(I, P(n,m), J(N))$ represents the rational sequence
	\begin{quote}
		$\left( I + \sum_{i=0}^{n} \J_{j=1}^n[P(i,j),J(n)]2^{-i}\right) _n$
	\end{quote}
	Where $\J_{j=1}^n[P(i,j),J(n)]$ means we take the $J(n)$th element (or the last element if there are fewer than $J(n)$ elements) of the subsequence $\{P(i,j=2\ldots n)\,|\,P(i,j)\neq P(i,j-1)\}$. That is we take the last value, or the value which represent the $J(n)$th change, whichever is first.
\end{definition}

Thus in the NGCNs, as the power series is expanded, coefficients that have already been calculated may change, provided there is a computable bound on how often they change. Here we do not need negative coefficients as we are able to correct coefficients.

Again, it should be clear that NGCN elements can be written in a formal syntax so it forms a decidable/computable set. Every NGCN represents a computable convergent rational sequence, albeit in general not modulated, i.e. we do cannot know how far in the sequence we need to advance to achieve a specific level of approximation. Thus NGCN is a Decidable class of (Unmodulated) Halting Computed Numbers.

Clearly the Specker sequence can be represented here with $J(n)=1$, so in that sense the GCNs can represent numbers which are not classical computable numbers, hence the name Generalized.

Unfortunately this definition gives too much weight to the binary expansions and does not conveniently support operations such as multiplication. An alternative approach may by provided by constraining a general program rather than constructing a power series.

\subsection{Constrained GCNs}
\begin{definition}
	A \textbf{Generalized Computed Number} (GCN) is a triple $(P, M, j)$ where $M \in \N$ is an integer bound on the output, $j$ is a primitive recursive function giving a bound on the number of jumps of any order of magnitude, and $P$ is a program which outputs rational numbers which we will run within a specific wrapper program $G$
	which ensures that 
	\begin{itemize}
		\item $P$ outputs at least one rational number (say it starts by outputting $0$) 
		\item Every output satisfies $|q_i| \leq M$ (if not the program will repeat the previous output instead)
		\item The $n$th output is a rational number $q_n$ such that for every $k \leq n$ there does not exist $j(k)$ non-overlapping pairs $1\leq a_1<b_1<a_2<b_2<\ldots <a_{j(k)}<b_{j(k)}\leq n$ with $|q_{a_i}-q_{b_i}|>2^{-n}$ (if not the program will repeat the previous output instead).
	\end{itemize}
	
\end{definition}

With this definition we have a very flexible Decidable Class of Generalized Halting Computed Numbers. 
They are however Unmodulated. 

$P$ is not Halting as a general computer program $P$ may get stuck in a loop.
If $P$ is limited to be primitive recursive algorithms, then the GCN is Halting in the weak sense that it will always output an infinite sequence, but the sequence may get stuck repeating the same number while $P$ outputs numbers that are censored by $G$.  

\subsection{GCN arithmetic operations}
\subsubsection{Addition}
Given two \textbf{GCN}s $X = (P_X, M_X, j_X)$ and 
$Y = (P_Y, M_Y, j_Y)$ we define the addition $X + Y = (P_{X+Y}, M_X+M_Y, J_X+J_Y)$ where $P_{X+Y}$ outputs zero and then executes $P_X$ and $P_Y$ in parallel (or alternatively executes a step of each); whenever a new rational number is output from $P_X$ (or $P_Y$ respectively) $P_{X+Y}$ outputs the new value added to the last known value from $P_Y$ ($P_X$) or zero if none.

\subsection{Additive inverse}
This is obtained trivially by $-X = (-P_X, M_X, j_X)$ where $-P_X$ is the program $P_X$ with an extra wrapper that negates each rational number coming out of $P_X$.

\subsection{Multiplication}
Given two \textbf{GCN}s $X = (P_X, M_X, j_X)$ and 
$Y = (P_Y, M_Y, j_Y)$ we define the multiplication $X*Y = (P_{X*Y}, M_X * M_Y, j_{X*Y})$ as follows:
 \begin{itemize}
  \item 
$P_{X*Y}$ is a computer program which wraps $P_X$ and $P_Y$ and outputs zero and then executes $P_X$ and $P_Y$ in parallel (or alternately executes a step of each); whenever a new rational number is output from $P_X$ (or $P_Y$) $P_{X*Y}$ outputs the new value multiplied by the most recent value from $P_Y$ ($P_X$) or zero if none.
  \item 
To construct $J_{X*Y}$ first pick the smallest $m_x$ and $m_y$ such that $M_X \leq 2^{m_x}$, $M_Y \leq 2^{m_y}$.
  \begin{itemize}
    \item
    $J_{X*Y}(1) = \sum_{1 \leq i \leq m_y}J_X(i) + \sum_{1 \leq i \leq m_x}J_Y(i)$
    \item
    $J_{X*Y}(k)=J_X(k+m_y)+J_Y(k+m_x) \,\, (k>1)$. 
 \end{itemize}
 \end{itemize}

We see here the need for having a bound $M$. Having a bound on the numbers being output by $X$ allows us to limit the impact of the jumps in $Y$ after multiplying by numbers in $X$, and vice versa.

\section{Conclusion}
We suggested an approach of epi-constructionism and explored computable sets of computable numbers, or, equivalently, decidable constructive real numbers.
We were able to define Computed Numbers as fundamental objects without relying on the abstraction of real numbers. The Computed Numbers are algorithms which output convergent sequences of rational numbers. This much was well known in the literature. Here we studied Decidable classes of Computed Numbers.

The best behaved class are the Signed Primitive Recursive Computed Numbers given by a binary power series with coefficients yielded by primitive recursive functions, allowing for negative coefficients. This means we can obtain an approximation of the limit to any accuracy in finite time. The class is closed under addition, subtracting and multiplication but there is no division and we cannot in general determine equality of the limit. 

A more general class, the Recursive Computed Numbers, corresponds to the concrete representations of the classical computable numbers, but this class is not closed under arithmetic operations and is not ``halting'' - the algorithm outputting a sequence might become stuck in an infinite loop. 

Finally we introduced the Generalized Computed Numbers which admit decidable representations of non-computable numbers such as the Specker limit, and still satisfy the requirement that each Generalized Computed Number has a concrete decidable syntactic representation. However these sequences are not modulated - we do not know when an output sequence will settle down to within a given error range. GCNs might be considered curiosities, they show that the concept of a computable set of computable numbers can stretch to include represent the Specker limit which is not classically a computable number. But GCNs are not a particularly natural set and being unmodulated, are not suitable for computation in practice.
 
It is the signed primitive recursive computed numbers which are the most likely foundation for a decidably constructive mathematics.

\section{Acknowledgments}
My thanks to Frank Waaldijk, Nachi Avraham Reem, Joshua Fox, Ziv Hellman, Benjamin Fox, Thorsten Altenkirch for their constructive feedback.

\section{Disclosure}
No funding was received. The author has no conflicts of interest.

\bibliographystyle{alpha}
\bibliography{myrefs}  

\begin{thebibliography}{{Tur}36}

\bibitem[Abe01]{aberth2001computable}
Oliver Aberth.
\newblock {\em Computable Calculus}.
\newblock Elsevier Science, 2001.

\bibitem[Bis67]{bishop1967foundations}
Errett Bishop.
\newblock {\em Foundations of constructive analysis}.
\newblock McGraw-Hill series in higher mathematics. McGraw-Hill, 1967.

\bibitem[Bri06]{Bridger2006}
Mark Bridger.
\newblock {\em Real Analysis: A Constructive Approach 1st Edition}.
\newblock Wiley-Interscience, 2006.

\bibitem[Bri19]{BridgesVita2019}
Mark Bridger.
\newblock {\em Real Analysis: A Constructive Approach Through Interval
  Arithmetic}.
\newblock American Math Society, 2019.

\bibitem[Bro23]{Brouwer1923}
Luitzen E.~J. Brouwer.
\newblock {Über Definitionsbereiche von Funktionen} [on the significance of
  the principle of excluded middle in mathematics, especially in function
  theory].
\newblock {\em Mathematische Annalen}, 97:60–75, 1923.

\bibitem[Chu36]{Church1936b}
Alonzo Church.
\newblock A note on the entscheidungsproblem.
\newblock {\em J. Symbolic Logic}, 1(1):40--41, 03 1936.

\bibitem[Coh66]{Cohen1966}
Paul~Joseph Cohen.
\newblock {\em Set theory and the continuum hypothesis}.
\newblock Dover Publications, 1966.

\bibitem[CSH08]{hardin2008}
A.~D.~Taylor C.~S.~Hardin.
\newblock {An.lntroduction to Infinite Hat Problems}.
\newblock {\em Springer Science+Business Media}, 30(4), 2008.

\bibitem[Dav58]{Davis1982}
Martin Davis.
\newblock {\em Computability \& Unsolvability}.
\newblock McGraw-Hill, 1958.

\bibitem[DSB06]{BridgesVita2006}
Luminita Simona~Vita Douglas S.~Bridges.
\newblock {\em Techniques of Constructive Analysis}.
\newblock 2006.

\bibitem[Gol42]{Goldbach1742}
Christian Goldbach.
\newblock letter to {E}uler.
\newblock 1742.

\bibitem[Hat92]{hata1992}
Masayoshi Hata.
\newblock Improvement in the irrationality measures of $\pi$ and $\pi^2$.
\newblock {\em Proc. Japan Acad. Ser. A Math. Sci.}, 68(9):283--286, 1992.

\bibitem[Kah04]{Kahan2004}
William Kahan.
\newblock A logarithm too clever by half.
\newblock 2004.

\bibitem[Kus06]{Kushner2006}
Boris~A. Kushner.
\newblock The constructive mathematics of {A}. {A}. {M}arkov.
\newblock {\em The American Mathematical Monthly}, 113(6):559--566, 2006.

\bibitem[LMT98]{Lefevre1998}
Vincent Lefèvre, Jean-Michel Muller, and Arnaud Tisserand.
\newblock Toward correctly rounded transcendentals.
\newblock {\em IEEE Transactions on Computers}, 47:1235 -- 1243, November 1998.

\bibitem[Mah53]{KMahler1954}
Kurt Mahler.
\newblock On the approximation of $\pi$.
\newblock 1953.

\bibitem[Mar62]{Markov1962}
Andrey~A. Markov.
\newblock On constructive mathematics.
\newblock {\em Trudy Mat. Inst. Steklov, Problems of the constructive direction
  in mathematics. Part 2.}, 67:8--14, 1962.

\bibitem[Mar71]{Markov1971translation}
Andrey~A. Markov.
\newblock On constructive mathematics (translation from russian paper of 1962.
\newblock {\em Amer. Math. Soc. Translations}, 2(98), 1971.

\bibitem[Myh53]{myhill1953}
John Myhill.
\newblock Criteria of constructibility for real numbers.
\newblock {\em J. Symbolic Logic}, 18(1):7--10, 03 1953.

\bibitem[Pea89]{peano1889arithmetices}
G.~Peano.
\newblock {\em Arithmetices principia: nova methodo}.
\newblock Fratres Bocca, 1889.

\bibitem[Ric54]{rice1954recursive}
H~Gordon Rice.
\newblock Recursive real numbers.
\newblock {\em Proceedings of the American Mathematical Society},
  5(5):784--791, 1954.

\bibitem[Rob51]{robinson1951}
R.~M. Robinson.
\newblock Review of {“R. Peter: Rekursive Funktionen”}.
\newblock {\em J. Symbolic Logic}, 16:280 – 282, 1951.

\bibitem[RZ06]{RETTINGER2006818}
Robert Rettinger and Xizhong Zheng.
\newblock A hierarchy of turing degrees of divergence bounded computable real
  numbers.
\newblock {\em Journal of Complexity}, 22(6):818 -- 826, 2006.
\newblock Computability and Complexity in Analysis.

\bibitem[Sha55]{rice1955}
Norman Shapiro.
\newblock H. g. rice. recursive real numbers. proceedings of the american
  mathematical society, vol. 5 (1954), pp. 784–791.
\newblock {\em Journal of Symbolic Logic}, 20(2):177–177, 1955.

\bibitem[Spe49]{Specker49}
Ernst Specker.
\newblock Nicht konstruktiv beweisbare {S}{\"a}tze der {A}nalysis
  [nonconstructive theorems of analysis].
\newblock {\em J. Symb. Log.}, 14(3):145--158, 1949.

\bibitem[{Tur}36]{TURING1936}
Alan~M. {Turing}.
\newblock {On computable numbers, with an application to the
  Entscheidungsproblem.}
\newblock {\em {Proc. Lond. Math. Soc. (2)}}, 42:230--265, 1936.

\bibitem[Tur37]{Turing1937}
Alan~M. Turing.
\newblock {On Computable Numbers, with an Application to the
  Entscheidungsproblem}.
\newblock {\em Proceedings of the London Mathematical Society},
  s2-42(1):230--265, 01 1937.

\bibitem[Wei00]{Weihrauch2000}
Klaus Weihrauch.
\newblock {\em Computable Analysis: An Introduction}.
\newblock Springer-Verlag, Berlin, Heidelberg, 2000.

\bibitem[ZR04]{ZhengRettinger2004}
Xizhong Zheng and Robert Rettinger.
\newblock Weak computability and representation of reals.
\newblock {\em Mathematical Logic Quarterly}, 50(0):431--442, 2004.

\end{thebibliography}

\end{document}